\documentclass{article}
\usepackage{graphicx}  
\usepackage{amsmath}   
\usepackage[compress,numbers,sort]{natbib}
\usepackage{amssymb}   
\usepackage{bm} 
\usepackage{dcolumn}
\usepackage{color}
\usepackage{mathrsfs}
\usepackage{amsfonts}
\usepackage{varioref}
\usepackage{textcomp}
\usepackage[normalem]{ulem}

\usepackage{multirow}
\usepackage{caption}
\usepackage{subcaption}
\RequirePackage[colorlinks,citecolor=blue,urlcolor=magenta,linkcolor=blue]{hyperref}
\allowdisplaybreaks
\labelformat{section}{Section #1} 
\labelformat{subsection}{Section #1} 
\labelformat{subsubsection}{Section #1}
\labelformat{subsubsubsection}{Section #1}
\labelformat{equation}{Eq.~(#1)} 
\labelformat{figure}{Fig.~#1} 
\labelformat{subfigure}{Fig.~\thefigure#1} 
\labelformat{table}{Table~#1} 
\labelformat{appendix}{Appendix #1}
\title{\bf Scalar tidal response of a rotating BTZ black hole}
\author{Rajendra Prasad Bhatt\href{https://orcid.org/0009-0004-9088-2998}{\includegraphics[scale=0.07]{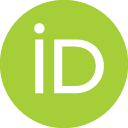}}\footnote{\href{mailto:rajendra@iucaa.in}{rajendra@iucaa.in}} ~and Chiranjeeb Singha\href{https://orcid.org/0000-0003-0441-318X}{\includegraphics[scale=0.07]{ORCIDiD_icon128x128.png}}\footnote{\href{mailto:chiranjeeb.singha@iucaa.in}{chiranjeeb.singha@iucaa.in}}\\
{\small{Inter-University Centre for Astronomy and Astrophysics, Pune 411007, India}}\\
}
\begin{document}
  
\maketitle
\begin{abstract} 

We study the response of a rotating BTZ black hole to the scalar tidal perturbation. We show that the real component of the tidal response function isn't zero, indicating that a rotating BTZ black hole possesses non-zero tidal Love numbers. Additionally, we observe scale-dependent behaviour, known as log-running, in the tidal response function. We also conduct a separate analysis on an extremal rotating BTZ black hole, finding qualitative similarities with its non-extremal counterpart. In addition, we present a procedure to calculate the tidal response function of a charged rotating BTZ black hole as well.

\end{abstract}

\section{Introduction}

Einstein's general theory of relativity describes gravity as a manifestation of the spacetime curvature~\cite{Wald:1984rg, Chandrasekhar:1985kt}. It provides various exciting predictions. 
 One of them is a black hole. 
Various black hole solutions have been found in $3+1$ spacetime dimensions~\cite{Schwarzschild:1916uq, Kerr:1963ud,Wald:1984rg, Chandrasekhar:1985kt}.
Similar black hole solutions have also been explored in lower~\cite{Banados:1992wn, Banados:1992gq} and higher spacetime dimensions~\cite{Myers:1986un} than $3+1$. In this paper, we will study about a black hole in $2+1$ spacetime dimensions.

In 1992, M. Bañados, C. Teitelboim, and J. Zanelli showed that in $2+1$ spacetime dimensions, Einstein-Maxwell equation with negative cosmological constant gives a black hole solution~\cite{Banados:1992wn, Banados:1992gq} (for review see ref.~\cite{Carlip:1995qv}). These solutions are defined by their mass, angular momentum, and charge. Based on the name of the authors, this black hole is famously known as the BTZ black hole. While BTZ black hole has event and Cauchy horizons (for rotating, charge, and charged rotating cases), it does not have a curvature singularity at the origin~\cite{Carlip:1995qv}.
The spacetime metric of a rotating BTZ black hole is given by~\cite{Dias:2019ery},
\begin{equation}\label{BTZ_metric}
    \mathrm{d}s^2 = -f(r)\mathrm{d}t^2 + \frac{1}{f(r)}\mathrm{d}r^2 + r^2\left[\mathrm{d}\phi-\Omega(r)\mathrm{d}t\right]^2~,
\end{equation}
where
\begin{equation}\label{def_f_Omega}
    f(r) = \frac{(r^2-r_+^2)(r^2-r_-^2)}{l^2r^2}~,\qquad \Omega(r) = \frac{r_+r_-}{lr^2}~,
\end{equation}
and $r_-$ and $r_+$ are the radial positions of the Cauchy (inner) and the event (outer) horizons, respectively, defined as
\begin{equation}\label{horizons}
    \frac{2r_\pm^2}{l^2} = M\left(1\pm\sqrt{1-\frac{J^2}{M^2l^2}}\right)~.
\end{equation}
Here $M$ is the mass, $J$ is the angular momentum of the black hole, and $l$ is the anti-de Sitter (AdS) radius. The cosmological constant ($\Lambda$) for this black hole spacetime is negative and related to the AdS radius as $\Lambda\equiv-1/l^2$~\cite{Dias:2019ery, Rocha:2011wp}.

In this work, we will study the response of a rotating BTZ black hole to the scalar tidal field.
The tidal response of a self-gravitating object can be divided into two parts: conservative (tidal deformation) and dissipative (tidal dissipation)~\cite{Chia:2020yla}. One can associate dimensionless numbers with these parts, known as tidal Love numbers (TLNs) and tidal dissipation numbers, respectively~\cite{Chia:2020yla, poisson_will_2014}.\footnote{In the external observer's frame, the tidal field can be static (time-independent) or dynamic (time-dependent) in nature, thus, the tidal response function is called static or dynamical tidal response function, respectively (similarly, for tidal Love numbers and tidal dissipation numbers)~\cite{poisson_will_2014}.} Various studies suggest vanishing static TLNs for black holes in $3+1$ dimensional spacetime~\cite{Binnington:2009bb, Damour:2009vw, Kol:2011vg, Chakrabarti:2013lua, Gurlebeck:2015xpa, Landry:2015zfa, Pani:2015hfa, Chia:2020yla, Charalambous:2021mea, Hui:2020xxx, Creci:2021rkz, Pereniguez:2021xcj, Bhatt:2023zsy, Sharma:2024hlz, Rai:2024lho, Bhatt:2024yyz, Kehagias:2024yzn}
. In addition, refs.~\cite{Charalambous:2023jgq, Rodriguez:2023xjd} reported non-zero static scalar TLNs for a 5-dimensional Myers-Perry black hole. Furthermore, TLNs have been found to be non-zero for braneworld black holes (presence of an extra spatial dimension)~\cite{Chakravarti:2018vlt}.\footnote{Apart from that, TLNs have also been studied for ultra-compact objects~\cite{Cardoso:2017cfl,Chakraborty:2023zed}, black hole in the alternative theory of gravity~\cite{DeLuca:2022tkm}, black hole in Schwarzschild de Sitter spacetime~\cite{Nair:2024mya}, AdS black branes~\cite{Emparan:2017qxd}, and area-quantized black hole~\cite{Nair:2022xfm}.} Thus, the tidal response of black holes has been studied in $3+1$ and higher dimensions; it would be interesting to study the tidal response of a rotating BTZ black hole, which is in $2+1$ dimensional spacetime, and consequently analyze the nature of its TLNs. In this work, we will consider a scalar tidal field and study the response of a BTZ black hole to it.\footnote{Recently, a paper appeared on the arXiv preprint server~\cite{DeLuca:2024ufn}, which also presented the calculation of the scalar tidal response function of a rotating BTZ black hole. There are differences in the procedure adopted in our work and in ref.~\cite{DeLuca:2024ufn}. In addition, we have also presented a procedure to calculate the tidal response function of a charged rotating BTZ black hole.} These results might also be helpful to understand the hidden symmetries of general relativity~\cite{Charalambous:2021kcz, Charalambous:2022rre}.

This paper is arranged as follows: In the next section (\ref{sec_2}), we will briefly discuss the definition of the tidal response function. In \ref{sec_3}, we shall describe the calculation of the scalar tidal response function of a non-extremal rotating BTZ black hole. In \ref{sec_4}, we will present a similar calculation for an extremal rotating BTZ black hole. In \ref{sec_5}, we will propose a procedure to calculate the scalar tidal response function of a charged rotating BTZ black hole. Finally, we conclude the paper with a discussion of the results and potential future works. Some calculations are relegated to \ref{app_1} and \ref{app_2}.

{\em Notations and conventions}: 
In this paper, we will use a positive signature convention, i.e., the Minkowski metric for $2+1$ dimensions
in the Cartesian coordinate system is given by $\text{diag.}(-1, +1, +1)$.
We will set $G=c=1$ throughout the paper. $f'(x)$ will denote the first-order derivative of $f(x)$ with respect to $x$. 
\section{Definition of the tidal response function}\label{sec_2}

In this section, we will briefly review the definition of the tidal response function of a self-gravitating body (for more details, see~~\cite{poisson_will_2014, LeTiec:2020bos, Bhatt:2023zsy}). We will also describe the procedure to calculate the scalar tidal response function of a rotating BTZ black hole.

Let us consider a non-rotating 
self-gravitating body of mass $M$ in an external tidal field. In Fourier space, the gravitational potential at radial coordinate $r$ is given by~\cite{poisson_will_2014, LeTiec:2020bos, Chia:2020yla, Bhatt:2023zsy}
\begin{equation}\label{eq_4}
U = \frac{M}{r} - \sum_{l=2}^{\infty}\sum_{m=-l}^{l}\frac{(l-2)!}{l!}\mathcal{E}_{lm}r^l\left[1+F_{lm}(\omega)\left(\frac{\mathcal{R}}{r}\right)^{2l+1}\right]Y_{lm}~,
\end{equation}
where $M/r$ is a point particle contribution to the potential, $\mathcal{E}_{lm}$ are tidal moments, and $F_{lm}$ is defined as the response of the self-gravitating body to the tidal field~(or tidal response function; for detailed description, see~~\cite{poisson_will_2014, LeTiec:2020bos, Chia:2020yla, Bhatt:2023zsy}). In linear perturbation theory, it can be expressed as
$F_{lm} = 2k_{lm}  +i\omega\tau_0 \nu_{lm} + \mathcal{O}(\omega^{2})$, where $k_{lm}$ and $\nu_{lm}$ are the tidal Love numbers and tidal dissipation numbers, respectively.\footnote{Since we will be working in the $2+1$ spacetime dimension, only the azimuthal index ($m$) will be needed in our analysis. Hence, we will denote the response function as $F_m$ instead of $F_{lm}$ (similarly $k_{lm}\to k_m$ and $\nu_{lm} \to \nu_m$). Also, $l$ will only denote the AdS radius onwards.}
$\mathcal{R}$ is related to the radial scale of the body and $\tau_0$ is the viscosity induced time delay. 
Therefore, the ratio of the coefficients of the decaying part (apart from the point particle contribution) and the growing part of the potential is proportional to the tidal response function.

For gravitational perturbation, it is well known that the Newman-Penrose scalars are related to the gravitational potential in the Newtonian regime, and one can calculate the tidal response function from them~\cite{LeTiec:2020bos, Bhatt:2023zsy}.
In a similar way, we can calculate the response of a body to a scalar tidal field using the scalar field quantity (say $\phi$)~\cite{Teukolsky:1972my, Rodriguez:2023xjd} and define the tidal response function as follows~\cite{Creci:2021rkz}: {\em The ratio of the coefficients of the decaying part and the growing part of the field quantity $(\phi)$ is proportional to the tidal response function.}

Thus, our methodology for this work will be as follows: First, we shall calculate the form of the scalar perturbation field quantity $(\phi)$ from the source-free Klein-Gordon field equation. Since the Klein-Gordon field equation is a second-order differential equation, we will arrive at two linearly independent solutions. Then we will simplify the general solution by using the purely ingoing boundary condition at the event horizon. Finally, we will find out the behaviour of the field quantity $(\phi)$ in the intermediate region\footnote{By intermediate region, we mean the region far from the black hole, but well within the tidal field~\cite{Bhatt:2024yyz}. Thus, the intermediate region is defined as a region of large $r$ ($r\gg r_+$) such that $(r-r_+)~\omega\ll 1$, where $r$ is a radial distance from the central object (BTZ black hole in this paper) and $\omega$ is the mode frequency of the tidal field.} and identify the scalar tidal response function from the above-given definition.

In this work, we are interested in calculating the tidal response function of a BTZ black hole, whose spacetime is not asymptotically flat. However, the above stated definition to calculate the tidal response function is used for an asymptotically flat spacetime. Thus, a question arises: can we use the above written definition for a non-asymptotically flat spacetime? The answer is yes, because the calculation of the tidal response depends on the growing and decaying behaviour (power law behaviour) of the field quantity, not on the asymptotic of the spacetime~\cite{Nair:2024mya}.
Therefore, we can use this approach for a non-asymptotically flat spacetime. One can also use the worldline effective field theory and the scattering amplitude calculation to find the tidal response function of a BTZ black hole~\cite{Creci:2021rkz, Nair:2024mya}.

In the next sections, we will calculate the scalar tidal response function for a non-extremal rotating BTZ black hole and an extremal rotating BTZ black hole, respectively. We will also propose a procedure to calculate the scalar tidal response of a charge rotating BTZ black hole.

\section{Scalar tidal response function of a non-extremal rotating BTZ black hole}\label{sec_3}

In this section, we will calculate the scalar tidal response function of a non-extremal rotating BTZ black hole.
To calculate the response of an object to a tidal field, one needs to study perturbations on it. Since we are interested in the scalar perturbation, we will start our analysis with a source-free Klein-Gordon equation:
\begin{equation}\label{K_G}
    \frac{1}{\sqrt{-g}}\partial_\mu\left[\sqrt{-g}g^{\mu\nu}\partial_\mu\phi\right] =0~,
\end{equation}
where $\phi$ is the field quantity. Since we are working in $2+1$ dimension, let us consider a decomposition $\phi = R(r)e^{-i\omega t}e^{im\phi}$. Now, the radial part of the source-free Klein-Gordon equation satisfies
\begin{equation}\label{radial_perturbation}
    R''(r)+ \left[\frac{1}{r} + \frac{f'(r)}{f(r)}\right] R'(r) +\frac{1}{f(r)}\left[-\frac{m^2}{r^2}+\frac{1}{f(r)}\{\omega-m\Omega(r)\}^2\right]R(r)=0~.
\end{equation}
We can simplify this equation further by considering a coordinate transformation~\cite{Dias:2019ery}
\begin{equation}\label{transformation}
    z=\frac{r^2-r_+^2}{r_+^2-r_-^2}~,
\end{equation}
which shifts the origin at the event horizon. However, the above-defined coordinate transformation is not applicable for the extremal rotating BTZ black hole because it diverges for the extremal rotating BTZ black hole ($r_+ = r_-$). Therefore, the calculation performed in this section is only valid for the non-extremal rotating BTZ black hole. In the next section, we will study the tidal response of an extremal rotating BTZ black hole separately.

The radial source-free Klein-Gordon equation transforms to (for elaborate calculation, see \ref{app_1a})
\begin{equation}\label{transformed_K_G_eq}
    R''(z) + \left[\frac{1}{z}+\frac{1}{1+z}\right]R'(z)+\frac{1}{z^2(1+z)^2}\left[\left\{\frac{(\omega-m\Omega_+)^2}{4\kappa_+^2}-\frac{(\omega-m\Omega_-)^2}{4\kappa_-^2}\right\}z +\frac{(\omega-m\Omega_+)^2}{4\kappa_+^2}\right]R(z)=0~,
\end{equation}
where $\kappa_\pm$ and $\Omega_\pm$ are the (positive) surface gravities and the angular velocities corresponding to the event and Cauchy horizons, respectively~\cite{Dias:2019ery}: 
\begin{equation}\label{def_kappa_Omega_pm}
    \kappa_\pm = \frac{r_+^2-r_-^2}{l^2r_\pm}~, \qquad \Omega_\pm = \frac{r_\mp}{lr_\pm}~.
\end{equation}
The above equation can be exactly solved in terms of Gauss hypergeometric functions. The solution is given by~\cite{Dias:2019ery} (for detailed analysis, see \ref{app_1b})
\begin{multline}\label{solution_1}
    R(z) = C_1z^{-i\frac{\omega-m\Omega_+}{2\kappa_+}}(1+z)^{-i\frac{\omega-m\Omega_-}{2\kappa_-}} \,_{2}F_{1}(a,b;c;-z) \\+ C_2 z^{i\frac{\omega-m\Omega_+}{2\kappa_+}} (1+z)^{-i\frac{\omega-m\Omega_-}{2\kappa_-}}\,_{2}F_{1}(a+1-c,b+1-c;2-c;-z)~,
\end{multline}
where $C_1$ and $C_2$ are the constant of integration and $a$, $b$, and $c$ are:
\begin{equation}\label{def_a_b_c}
    a= \frac{1}{2}\left(- i\frac{\omega-m\Omega_+}{\kappa_+} - i\frac{\omega-m\Omega_-}{\kappa_-}\right)~,\qquad b= \frac{1}{2}\left(2 - i\frac{\omega-m\Omega_+}{\kappa_+} - i\frac{\omega-m\Omega_-}{\kappa_-}\right)~,\qquad c = 1 - i\frac{\omega-m\Omega_+}{\kappa_+}~.
\end{equation}
Now we can put a constraint on the constants of integration by using the ingoing boundary condition at the event horizon. Near the event horizon ($r=r_+$), i.e., $z\rightarrow0$, the solution becomes
\begin{equation}
    R(z) \sim C_1 z^{-i\frac{\omega-m\Omega_+}{2\kappa_+}}(1+z)^{-i\frac{\omega-m\Omega_-}{2\kappa_-}} + C_2 z^{i\frac{\omega-m\Omega_+}{2\kappa_+}} (1+z)^{-i\frac{\omega-m\Omega_-}{2\kappa_-}}~.
\end{equation}
Based on the decomposition $(\phi = R(r)e^{-i\omega t}e^{im\phi})$, the second part of the above equation is outgoing in nature. Therefore, for a solution to be purely ingoing at the event horizon, $C_2$ must be zero. Now, the solution becomes
\begin{equation}
    R(z) = C_1 z^{-i\frac{\omega-m\Omega_+}{2\kappa_+}}(1+z)^{-i\frac{\omega-m\Omega_-}{2\kappa_-}} \,_{2}F_{1}(a,b;c;-z)~.
\end{equation}
To calculate the tidal response function, we need to calculate the large $z$ or large $r$ limit of the above solution (behaviour of the solution in the intermediate region), which will be
\begin{equation}\label{large_z_expansion}
    R(z) \sim \frac{z^{-a} \Gamma (c)}{\Gamma (a+1) \Gamma (c-a)}\left[1+z^{-1} a (1+a-c)\left\{1-\psi(c-a-1)-\psi(a+1)+\ln (z)-2\gamma\right\}\right]~,
\end{equation}
where $\psi(x)$ is a digamma function defined as $\Gamma'(x)/\Gamma(x)$, and $\gamma$ is Euler-Mascheroni constant having value $0.577216$~\cite{Arfken_Weber_2005}.
Now, following the definition given in \ref{sec_2}, we can identify the scalar tidal response function as
\begin{equation}\label{resp_func_1}
    F_m = a (1+a-c)\left[1-\psi(c-a-1)-\psi(a+1)+\ln (z) -2\gamma \right]~.
\end{equation}
The response function depends on the radial coordinate due to the $\ln(z)$ term, which can be termed as the log-running (scale dependent) behaviour of the tidal response function~\cite{Katagiri:2023umb, DeLuca:2022tkm}. We can define the tidal response function independent of the radial coordinate as well:\footnote{Where we have rewritten the \ref{large_z_expansion} as
\begin{equation}
    R(z) \sim \frac{z^{-a} \Gamma (c)}{\Gamma (a+1) \Gamma (c-a)}\left[1+ \Tilde{F}_m z^{-1} \left\{1+\frac{1}{\Tilde{F}_m}\ln(z)\right\}\right]~,
\end{equation}
and defined the tidal response function as the coefficient of $z^{-1} \left\{1+(\ln(z)/\Tilde{F}_m)\right\}$ in the square bracket.}
\begin{equation}\label{resp_func_1_non-running}
    \Tilde{F}_m = a (1+a-c)\left[1-\psi(c-a-1)-\psi(a+1) -2\gamma \right]~.
\end{equation}
The multiplicative factor $a(1+a-c)$ is
\begin{equation}\label{mult_a_and_1_p_a_m_c_value}
    a(1+a-c) = \frac{\omega^2l^4-m^2l^2}{4(r_+^2-r_-^2)}~,
\end{equation}
which is a real quantity. Therefore, the tidal response function will have non-zero real parts. Since TLNs are related to the real part of the tidal response function, it implies that the scalar TLNs of a non-extremal rotating BTZ black hole are generically non-zero, and it can be expressed as
\begin{equation}
    k_m = \frac{1}{2} \text{Re}(\Tilde{F}_m) \equiv \frac{\omega^2l^4-m^2l^2}{8(r_+^2-r_-^2)}\left[1-\text{Re}\left\{\psi(c-a-1)+\psi(a+1)\right\} -2\gamma \right]~.
\end{equation}
Similarly, we can also write the expression of the scalar TLNs corresponding to the tidal response function with log-running behaviour (\ref{resp_func_1}).

Thus, the scalar tidal response function and scalar TLNs of a non-extremal rotating BTZ black hole are non-vanishing in general, but they vanish in some cases:
\begin{itemize}
    \item The tidal response function vanishes for $\omega = 0 = m$. It implies that the static TLNs of a non-extremal rotating BTZ black hole vanish in an axi-symmetric scalar tidal field. However, they are non-zero in general.
    \item Tidal response function also vanishes for $l^2\omega^2 = m^2$. The above-stated condition is a specific case of this condition.
\end{itemize}
For a {\em non-rotating BTZ black hole} ($\Omega(r)=0$), the final form of the response function is the same as \ref{resp_func_1} (or \ref{resp_func_1_non-running}), but the expressions of $a$ and $c$ will be different:
\begin{equation}
    a = -\frac{i(l^2\omega - ml)}{2r_+}~, \qquad c = 1-\frac{il^2\omega}{r_+}~.
\end{equation}
The multiplication factor for a non-rotating BTZ black hole will be $a(1+a-c) = (\omega^2l^4-m^2l^2)/4r_+^2$. Thus, the scalar TLNs of a non-rotating BTZ black hole are also non-zero. However, they vanish for the same cases listed above for the rotating BTZ black hole.

\section{Scalar tidal response function of an extremal rotating BTZ black hole}\label{sec_4}

In the last section, we have calculated the scalar tidal response function of an arbitrarily rotating BTZ black hole, and we found that it has a real part. However, the calculation performed in the last section is not valid for an extremal black hole $(r_+ = r_-)$ due to the transformation (\ref{transformation}), which diverges for an extremal rotating BTZ black hole. In this section, we will separately analyze the tidal response function for an extremal rotating BTZ black hole.

Since the transformed coordinate $z$ (\ref{transformation}) is not well defined for an extremal rotating BTZ black hole, we redefine the coordinate transformation
as
\begin{equation}\label{transformation_1}
    \tilde{z}=\frac{r^2-r_+^2}{r_+^2}~.
\end{equation}
Similar to the transformed coordinate in the last section (\ref{transformation}), this coordinate's origin is at the horizon. But contrary to \ref{transformation}, it is well defined for extremal rotating BTZ black hole. With the new coordinate, the radial source-free Klein-Gordon equation becomes (see \ref{app_2a})
\begin{equation}\label{eq_ext_BTZ_BH}
    R''(\tilde{z}) + \frac{2}{\tilde{z}}R'(\tilde{z})+\frac{1}{4\tilde{z}^4}\left(AB\tilde{z} +B^2\right)R(\tilde{z})=0~,
\end{equation}
where
\begin{equation}\label{def_A_B}
    A = \frac{l^2}{r_+}\left(\omega+\frac{m}{l}\right)~, \qquad B = \frac{l^2}{r_+}\left(\omega-\frac{m}{l}\right)~.
\end{equation}
The above equation can be exactly solved in terms of the confluent hypergeometric functions, and the solution is (see \ref{app_2b})
\begin{equation}\label{ext_BTZ_BH_sol}
    R(\tilde{z}) = C_1 e^{\frac{iB}{2\tilde{z}}} U\left(-\frac{iA}{4},0,-\frac{iB}{\tilde{z}}\right)+C_2 e^{-\frac{iB}{2\tilde{z}}} U\left(\frac{iA}{4},0,\frac{iB}{\tilde{z}}\right)~,
\end{equation}
where $C_1$ and $C_2$ are the constants of integration.
Now, we can apply the ingoing boundary condition at the event horizon to constrain the constants of integration. Near the event horizon ($r=r_+$), i.e., $\tilde{z}\rightarrow0$, the solution becomes
\begin{equation}
    R(\tilde{z}) \sim C_1 \left(-iB\right)^{\frac{iA}{4}}\left(\tilde{z}\right)^{-\frac{iA}{4}}e^{\frac{iB}{2\tilde{z}}} + C_2 \left(iB\right)^{-\frac{iA}{4}} \left(\tilde{z}\right)^{\frac{iA}{4}} e^{-\frac{iB}{2\tilde{z}}}~.
\end{equation}
Based on the decomposition $(\phi = R(r)e^{-i\omega t}e^{im\phi})$, for the solution to be purely ingoing at the event horizon, $C_2$ must be zero. Thus, the solution becomes
\begin{equation}
    R(\tilde{z}) = C_1 e^{\frac{iB}{2\tilde{z}}} U\left(-\frac{iA}{4},0,-\frac{iB}{\tilde{z}}\right).
\end{equation}
The large $\tilde{z}$ or large $r$ limit of the above solution (behaviour of the solution in the intermediate region) is
\begin{equation}\label{large_z_expansion_ext_BTZ}
    R(\tilde{z}) \sim \frac{1}{\tilde{a} \Gamma (\tilde{a})}\left[1 + iB\tilde{z}^{-1}\frac{ \left\{-2 \tilde{a} \ln \left(-iB\right) + 2 \tilde{a} \ln \left(\tilde{z}\right)-4 \gamma  \tilde{a}+2 \tilde{a}-2 \tilde{a} \psi(\tilde{a}+1)+1\right\}}{2}\right]~,
\end{equation}
where $\tilde{a}= -iA/4$, $\psi(x)$ is a digamma function defined as $\Gamma'(x)/\Gamma(x)$, and $\gamma$ is Euler-Mascheroni constant having value $0.577216$~\cite{Arfken_Weber_2005}.
Now, using the definition provided in \ref{sec_2}, we can identify the scalar tidal response function as
\begin{equation}\label{resp_func_2}
    F_m = \frac{B}{2}\left[i+\frac{A}{2}\left\{1-\ln \left(-iB\right)+\ln \left(\tilde{z}\right)-2 \gamma- \psi\left(1-\frac{iA}{4}\right)\right\} \right]~.
\end{equation}
Similar to the previous section, the tidal response function depends on $\ln(\tilde{z})$ (log-running behaviour of the tidal response function~\cite{Katagiri:2023umb, DeLuca:2022tkm}). We can also define the tidal response function as
\begin{equation}\label{resp_func_2_non_running}
    \Tilde{F}_m = \frac{B}{2}\left[i+\frac{A}{2}\left\{1-\ln \left(-iB\right)-2 \gamma- \psi\left(1-\frac{iA}{4}\right)\right\} \right]~,
\end{equation}
which is independent of $\ln(\tilde{z})$.~\footnote{Here too, we have rewritten the \ref{large_z_expansion_ext_BTZ} as
\begin{equation}
    R(\tilde{z}) \sim \frac{1}{\tilde{a} \Gamma (\tilde{a})}\left[1 + \Tilde{F}_m\tilde{z}^{-1}\left\{1+\frac{2\tilde{a}}{\Tilde{F}_m}\ln(\tilde{z})\right\}\right]~,
\end{equation}
and defined the tidal response function as the coefficient of $\tilde{z}^{-1} \left\{1+(2\tilde{a}\ln(\tilde{z})/\Tilde{F}_m)\right\}$ in the square bracket.}

The multiplicative factor outside the square bracket in the response function ($B/2$) is a real quantity, and there are real terms inside the square bracket as well.
Therefore, the response function will have non-zero real parts, which consequently implies that the extremal rotating BTZ black hole has non-vanishing scalar TLNs. TLNs of an extremal rotating BTZ black hole to a scalar tidal field can be written as
\begin{equation}
    k_m = \frac{1}{2} \text{Re}(\Tilde{F}_m) \equiv \frac{l(l\omega-m)}{4r_+}\left[\frac{A}{2}-A\gamma+\frac{A}{2}\,\text{Re}\left\{-\ln \left(-iB\right)- \psi\left(1-\frac{iA}{4}\right)\right\} \right]~.
\end{equation}
Similarly, we can also express the TLNs corresponding to the tidal response function with log-running behaviour (\ref{resp_func_2}).

Hence, the scalar tidal response function and scalar TLNs of an extremal rotating BTZ black hole are generally non-zero, but they can be zero for some specific cases:
\begin{itemize}
    \item The tidal response function is zero for $\omega=0=m$. Thus, static TLNs of an extremal rotating BTZ black hole also vanish for an axi-symmetric tidal field. However, they are non-zero in general.
    \item Tidal response function also vanishes for $l\omega = m$. It already satisfies the condition of the first case.
\end{itemize}
Having calculated the TLNs for non-extremal and extremal rotating BTZ black holes, we would like to see whether the extremal limit of the TLNs for the non-extremal rotating BTZ black hole gives the TLNs for extremal rotating BTZ black hole or not. Since coordinate transformations involved in the calculations for non-extremal and extremal rotating BTZ black holes are different and the tidal Love numbers show log-running behavior (depends on $z$ logarithmically) in both cases, the extremal limit of the TLNs for non-extremal rotating BTZ black hole will not give the TLNs calculated for extremal rotating BTZ black hole.

\section{Scalar tidal response function of a charged rotating BTZ black hole}\label{sec_5}

In this section, we shall study the response of a charged rotating BTZ black hole to a scalar tidal field. We will also study it for a non-rotating charged BTZ black hole.

The spacetime metric of a charged rotating BTZ black hole has the same form as given in \ref{BTZ_metric}, but the functional form of $f(r)$ is different~\cite{Banados:1992wn, Martinez:1999qi, Hendi:2010px, Singha:2022bvr}:
\begin{equation}\label{form_of_f(r)}
    f(r) = -M + \frac{r^2}{l^2} + \frac{J^2}{4r^2}- 2q^2\ln{\left(\frac{r}{l}\right)}~,\qquad \Omega(r) = \frac{J}{2r^2}~,
\end{equation}
where $q$ is the charge of the black hole.

To study the tidal response of a charged rotating BTZ black hole to the scalar tidal field, we study its perturbation using the Klein-Gordon equation. The radial part of the Klein-Gordon equation is given by \ref{radial_perturbation}:
\begin{equation}\label{source_fre_K_G_equation}
    R''(r)+ \left[\frac{1}{r} + \frac{f'(r)}{f(r)}\right] R'(r) +\frac{1}{f(r)}\left[-\frac{m^2}{r^2}+\frac{1}{f(r)}\{\omega-m\Omega(r)\}^2\right]R(r)=0~.
\end{equation}
For a charged rotating BTZ black hole, due to the logarithmic term in $f(r)$, we can't write $f(r)$ and $\Omega(r)$ as \ref{def_f_Omega}, which is
\begin{equation}
    f(r) = \frac{(r^2-r_+^2)(r^2-r_-^2)}{l^2r^2}~,\qquad \Omega(r) = \frac{r_+r_-}{lr^2}~,
\end{equation}
where $r_+$ and $r_-$ are the positions of the event and Cauchy horizons, respectively. Thus, we can't solve the radial Klein-Gordon equation by defining a new coordinate $z = (r^2-r_+^2)/(r_+^2-r_-^2)$, as done in \ref{sec_3}.

To proceed further, we will solve the radial Klein-Gordon equation for a slowly varying tidal field ($M\omega\ll 1$), and assume the charge ($q$) being very small, such that $q\sim\mathcal{O}(M\omega)$. Now, we can solve the Klein-Gordon equation in the near-zone region $((r-r_+)~\omega \ll 1)$
and calculate the tidal response function from it. This approach has been followed in various other studies of the tidal response of black holes~\cite{Chia:2020yla, Bhatt:2023zsy, Bhatt:2024yyz}. Let us consider a coordinate transformation $\tilde{z} = (r^2-r_+^2)/r_+^2$. In this new coordinate system, the event horizon is located at $\tilde{z}=0$. Now, we can write $f(r)$ as (see \ref{app_C} for intermediate steps)
\begin{equation}
    f(r) \sim -\tilde{M} + \frac{r^2}{\tilde{l}^2} +\frac{J^2}{4r^2}~,
\end{equation}
where we have neglected terms of order $q^2\mathcal{O}(\tilde{z}^2)$ or $\mathcal{O}(M^2\omega^2\tilde{z}^2)$ from $- 2q^2\ln{\left(r/l\right)}$ in $f(r)$ in the intermediate steps. Here, $\tilde{M}$ and $\tilde{l}$ are given by
\begin{equation}\label{tilde_M_tilde_l}
    \tilde{M} = M -q^2+q^2\ln{\left(\frac{r_+^2}{l^2}\right)}~, \qquad \tilde{l}^2 = l^2\left(1-\frac{q^2l^2}{r_+^2}\right)^{-1}~.
\end{equation}
Now we can express $f(r)$ and $\Omega(r)$ as \ref{def_f_Omega}:
\begin{equation}\label{charges_rotating_BTZ_f(r)_approximated}
    f(r) = \frac{(r^2-\tilde{r}_+^2)(r^2-\tilde{r}_-^2)}{\tilde{l}^2r^2}~, \qquad \Omega(r) = \frac{\tilde{r}_+\tilde{r}_-}{\tilde{l}r^2}~,
\end{equation}
with 
\begin{equation}
    \frac{2\tilde{r}_\pm^2}{\tilde{l}^2} = \tilde{M}\left(1\pm\sqrt{1-\frac{J^2}{\tilde{M}^2\tilde{l}^2}}\right)~.
\end{equation}
Since the approximated expressions of $f(r)$ and $\Omega(r)$ have the same form as \ref{def_f_Omega}, we can solve the source-free radial Klein-Gordon equation (\ref{source_fre_K_G_equation}) with the procedure followed in \ref{sec_3}. We can define a new coordinate $\bar{z} = (r^2-\tilde{r}_+^2)/(\tilde{r}_+^2-\tilde{r}_-^2)$,
which will transform \ref{source_fre_K_G_equation} to
\begin{equation}\label{transformed_K_G_eq_charged}
    R''(\bar{z}) + \left[\frac{1}{\bar{z}}+\frac{1}{1+\bar{z}}\right]R'(\bar{z})+\frac{1}{\bar{z}^2(1+\bar{z})^2}\left[\left\{\frac{(\omega-m\tilde{\Omega}_+)^2}{4\tilde{\kappa}_+^2}-\frac{(\omega-m\tilde{\Omega}_-)^2}{4\tilde{\kappa}_-^2}\right\}\bar{z} +\frac{(\omega-m\tilde{\Omega}_+)^2}{4\tilde{\kappa}_+^2}\right]R(\bar{z})=0~,
\end{equation}
where $\tilde{\kappa}_\pm$ and $\tilde{\Omega}_\pm$ are: 
\begin{equation}\label{def_kappa_Omega_pm_charged}
    \tilde{\kappa}_\pm = \frac{\tilde{r}_+^2-\tilde{r}_-^2}{\tilde{l}^2\tilde{r}_\pm}~, \qquad \tilde{\Omega}_\pm = \frac{\tilde{r}_\mp}{\tilde{l}\tilde{r}_\pm}~.
\end{equation}
The transformed source-free Klein-Gordon equation is same as \ref{transformed_K_G_eq} with $r_\pm$, $M$, and $l$ being replaced by $\tilde{r}_\pm$, $\tilde{M}$, and $\tilde{l}$. Thus, we can solve the above-written equation by following the procedure described in \ref{sec_3}. 
Assuming $\tilde{r}_+ \sim r_+$, the solution of the \ref{transformed_K_G_eq_charged} will be the same as given in \ref{sec_3}, with some appropriate changes given for the differential equation: $r_\pm$, $M$, and $l$ will be replaced by $\tilde{r}_\pm$, $\tilde{M}$, and $\tilde{l}$, respectively. Thus, the forms of the scalar tidal response function and the scalar TLNs for a charged rotating BTZ black hole, in the regime of very small charges, will also be the same as given in \ref{sec_3}, with the above-defined replacements $(\{r_{\pm}, \, M,\, l\} \to \{\tilde{r}_\pm,\, \tilde{M},\, \tilde{l}\})$.

Since the above defined coordinate $\bar{z}$ diverges for $\tilde{r}_+ = \tilde{r}_-$, we can perform a similar calculation given in \ref{sec_4} for this specific case. The results will be the same as given in \ref{sec_4} with the above-defined replacements $(\{r_\pm,\, M,\, l\} \to \{\tilde{r}_\pm,\, \tilde{M},\, \tilde{l}\})$.

The above results also apply to a charged non-rotating BTZ black hole if we set $J=0$.

\section{Discussion}\label{sec_6}

In this work, we studied the tidal response of a rotating (non-extremal and extremal) BTZ black hole to the scalar tidal perturbation. We found that the scalar tidal response function of a non-extremal rotating BTZ black hole is non-zero and has a real part, consequently implying non-zero TLNs. However, the tidal response function also vanishes in some cases (listed in \ref{sec_3}). Furthermore, static TLNs of a non-extremal rotating BTZ black hole to a scalar tidal field are generally non-vanishing. However, they vanish for an axi-symmetric tidal perturbation.

Since the transformed coordinate defined for the calculation of the scalar tidal response function of a non-extremal rotating BTZ black hole (\ref{transformation})
diverges for an extremal rotating BTZ black hole $(r_+=r_-)$, the calculation presented in \ref{sec_3} is not valid for it. Therefore, we separately calculated the scalar tidal response function for an extremal rotating BTZ black hole with a redefined coordinate transformation equation (\ref{transformation_1}). 
Qualitatively, the tidal response function of an extremal rotating BTZ black hole shows similar behaviour to that of a non-extremal rotating BTZ black hole. It is non-zero and has a real part, which implies non-zero TLNs. Here too, the tidal response function vanishes for certain cases (listed in \ref{sec_4}). In addition, static TLNs of an extremal rotating BTZ black hole to the scalar tidal perturbation are generally non-zero. However, they vanish for an axi-symmetric tidal perturbation too.

In addition, the tidal response function shows a log-running behaviour (depends on the radial coordinate)~\cite{Katagiri:2023umb, DeLuca:2022tkm, Charalambous:2023jgq, Rodriguez:2023xjd} for non-extremal and extremal rotating BTZ black holes. 
However, we have also given the form of the tidal response function, which is independent of the radial coordinate~\cite{Katagiri:2023umb, Bhatt:2024yyz}.

Our results can be summarized as follows:
\begin{itemize}
    \item Static and dynamic tidal Love numbers of non-extremal and extremal rotating BTZ black holes in a scalar tidal field are generically non-zero.
    \item Static tidal Love numbers of non-extremal and extremal rotating BTZ black holes vanish for an axi-symmetric scalar tidal perturbation.
\end{itemize}
Finally, we also proposed a procedure to calculate the tidal response function for a charged rotating BTZ black hole. Due to the logarithmic term in the metric of the charged BTZ black hole, we couldn't calculate the tidal response function in a similar way as done in \ref{sec_3} and \ref{sec_4}. Therefore, we did the calculation for a slowly varying tidal field in the near-zone region for a BTZ black hole with a small charge. With this assumption, the results for a charged rotating BTZ black hole will be similar to the results found in \ref{sec_3} and \ref{sec_4} with some replacements defined in \ref{sec_5} $(\{r_\pm,\, M,\, l\} \to \{\tilde{r}_\pm,\, \tilde{M},\, \tilde{l}\})$.
The suggested procedure is also applicable for a charged, non-rotating BTZ black hole, if we set $J=0$.

In this work, we saw that a rotating BTZ black hole responds to a scalar tidal field, and its tidal response function is complex (real + imaginary) in nature. 
Therefore, one can extend this work for electromagnetic and gravitational perturbations~\cite{Rocha:2011wp} and calculate the tidal response of a BTZ black hole for these perturbations.  
One can also do these calculations using worldline effective field theory and scattering amplitudes~\cite{Creci:2021rkz, Nair:2024mya}.

\section*{Acknowledgements}
We would like to thank Sumanta Chakraborty for suggesting this problem, and for his help throughout the project. R. P. B. would also like to thank Sukanta Bose for his support.


\appendix
\labelformat{section}{Appendix #1} 
\labelformat{subsection}{Appendix #1}
\section{Calculations for non-extremal rotating BTZ black hole}\label{app_1}
In this appendix, we will describe the calculation involved in arriving at \ref{transformed_K_G_eq} from \ref{radial_perturbation}. We will also describe the steps involved in calculating the solution of \ref{transformed_K_G_eq}.
\subsection{Transforming the radial Klein-Gordon equation}\label{app_1a}

The radial Klein-Gordon equation for BTZ black hole is (\ref{radial_perturbation})
\begin{equation}\label{radial_perturbation_app_a_1}
    R''(r)+ \left[\frac{1}{r} + \frac{f'(r)}{f(r)}\right] R'(r) +\frac{1}{f(r)}\left[-\frac{m^2}{r^2}+\frac{1}{f(r)}\{\omega-m\Omega(r)\}^2\right]R(r)=0~,
\end{equation}
where $f(r)$ and $\Omega(r)$ are defined in \ref{def_f_Omega}.
If we apply the coordinate transformation (\ref{transformation}), which is $z = (r^2-r_+^2)/(r_+^2-r_-^2)$, then \ref{radial_perturbation_app_a_1} transforms to
\begin{equation}\label{equation_perturbation_1}
    R''(z) + \frac{r_+^2-r_-^2}{2r^2}\left[2+r\frac{f'(r)}{f(r)}\right]R'(z)+\frac{(r_+^2-r_-^2)^2}{4r^2f(r)}\left[-\frac{m^2}{r^2}+\frac{1}{f(r)}\{\omega-m\Omega(r)\}^2\right]R(z)=0~.
\end{equation}
Now we can simplify the coefficient of $R'(z)$ and $R(z)$ separately.
Since
\begin{equation}
    \frac{f'(r)}{f(r)} = \frac{2r}{r^2-r_-^2} + \frac{2r}{r^2-r_+^2}-\frac{2}{r}~,
\end{equation}
the coefficient of $R'(z)$ can be reexpressed as:
\begin{equation}\label{coeff_R'(z)}
    \frac{r_+^2-r_-^2}{2r^2}\left[2+r\frac{f'(r)}{f(r)}\right] = \frac{1}{z}+\frac{1}{1+z}~.
\end{equation}
Similarly, the coefficient of $R(z)$ is:
\begin{equation}
    \frac{(r_+^2-r_-^2)^2}{4r^2f(r)}\left[-\frac{m^2}{r^2}+\frac{1}{f(r)}\{\omega-m\Omega(r)\}^2\right] = \frac{l^2}{4z(1+z)}\left[-\frac{m^2}{r^2}+\frac{1}{f(r)}\{\omega-m\Omega(r)\}^2\right]~.
\end{equation}
By using $r^2 = r_+^2 + (r_+^2-r_-^2)z$ and the form of $f(r)$ and $\Omega(r)$ from \ref{def_f_Omega}, the coefficient of $R(z)$ can be further simplified to
\begin{equation}
    \frac{1}{z^2(1+z)^2}\frac{l^4}{4(r_+^2-r_-^2)^2}\left[\left\{r_+^2\left(\omega-\frac{mr_-}{lr_+}\right)^2-r_-^2\left(\omega-\frac{mr_+}{lr_-}\right)^2\right\}z +r_+^2\left(\omega-\frac{mr_-}{lr_+}\right)^2\right]~.
\end{equation}
If we define $\kappa_\pm$ and $\Omega_\pm$ as in \ref{def_kappa_Omega_pm},
then the coefficient of $R(z)$ can be written as
\begin{equation}\label{coeff_R(z)}
    \frac{1}{z^2(1+z)^2}\left[\left\{\frac{(\omega-m\Omega_+)^2}{4\kappa_+^2}-\frac{(\omega-m\Omega_-)^2}{4\kappa_-^2}\right\}z +\frac{(\omega-m\Omega_+)^2}{4\kappa_+^2}\right]~.
\end{equation}
Therefore, \ref{equation_perturbation_1} becomes (using \ref{coeff_R'(z)} and \ref{coeff_R(z)})
\begin{equation}\label{equation_perturbation_1a}
    R''(z) + \left[\frac{1}{z}+\frac{1}{1+z}\right]R'(z)+\frac{1}{z^2(1+z)^2}\left[\left\{\frac{(\omega-m\Omega_+)^2}{4\kappa_+^2}-\frac{(\omega-m\Omega_-)^2}{4\kappa_-^2}\right\}z +\frac{(\omega-m\Omega_+)^2}{4\kappa_+^2}\right]R(z)=0~,
\end{equation}
which is our desired equation (\ref{transformed_K_G_eq}).

\subsection{Solution of the transformed radial Klein-Gordon equation}\label{app_1b}

In this section, we will describe the steps involved in arriving at the solution of \ref{transformed_K_G_eq}, which is
\begin{equation}\label{eq_2}
    R''(z) + \left[\frac{1}{z}+\frac{1}{1+z}\right]R'(z)+\frac{1}{z^2(1+z)^2}\left[P(1+z)-Qz \right]R(z)=0~,
\end{equation}
where 
\begin{equation}
    P = \frac{(\omega-m\Omega_+)^2}{4\kappa_+^2}~, \qquad Q = \frac{(\omega-m\Omega_-)^2}{4\kappa_-^2}~.
\end{equation}
If we make a transformation $x=-z$, then the above differential equation becomes
\begin{equation}\label{diff_eq_1}
    R''(x) + \left[\frac{1}{x}-\frac{1}{1-x}\right]R'(x)+\frac{1}{x^2(1-x)^2}\left[P(1-x)+Qx \right]R(x)=0~.
\end{equation}
It can be checked that this equation has three regular singular points at $x=0,\, 1,\text{ and } \infty$. The same property is also followed by the hypergeometric differential equation. Therefore, we can transform this equation to the hypergeometric differential equation with the substitution
\begin{equation}
    R(x) = x^\alpha (1-x)^\beta G(x)~,
\end{equation}
by choosing the suitable values of $\alpha$ and $\beta$. The above substitution transforms \ref{diff_eq_1} to
\begin{equation}\label{diff_eq_2}
    x(1-x)G''(x) +\left[1+2\alpha - 2x(1+\alpha+\beta)\right]G'(x)+ \left[-(\alpha+\beta)(1+\alpha+\beta)+\frac{P+\alpha^2}{x} + \frac{Q+\beta^2}{1-x}\right]G(x)=0~.
\end{equation}
For hypergeometric differential equation, there should not be $(P+\alpha^2)/x$ and $(Q+\beta^2)/(1-x)$ in the coefficient of $G(x)$. It means
\begin{equation}\label{eq_satisfy}
    P+\alpha^2 = 0~, \qquad Q+\beta^2 =0~.
\end{equation}
If we take any values of $\alpha$ and $\beta$ which satisfy \ref{eq_satisfy} (say $[\alpha_0, \beta_0]$), then \ref{diff_eq_2} becomes
\begin{equation}\label{equation_2}
    x(1-x)G''(x) +\left[1+2\alpha_0 - 2x(1+\alpha_0+\beta_0)\right]G'(x)-(\alpha_0+\beta_0)(1+\alpha_0+\beta_0)G(x)=0~.
\end{equation}
On comparing it with the hypergeometric differential equation~\cite{abramowitz_stegun_1972, Arfken_Weber_2005}
\begin{equation}
    x(1-x)\,G''(x) + [c-(a+b+1)x]\,G'(x)-ab\,G(x)=0~,
\end{equation}
which has the solution
\begin{equation}
    G(x) = \Tilde{C}_1 \,_{2}F_{1}(a,b;c;x) + \Tilde{C}_2\,x^{1-c}\,_{2}F_{1}(a+1-c,b+1-c;2-c;x)~,
\end{equation}
where $\Tilde{C}_1$ and $\Tilde{C}_2$ are the constant of integration,
we get
\begin{equation}
    a = \alpha_0+\beta_0~, \qquad  b = 1+\alpha_0+\beta_0~, \qquad c = 1+2\alpha_0~.
\end{equation}
Therefore, the solution of \ref{eq_2} is
\begin{equation}
    R(z) = (-z)^{\alpha_0}(1+z)^{\beta_0}\left[\Tilde{C}_1 \,_{2}F_{1}(a,b;c;-z) + \Tilde{C}_2\,(-z)^{1-c}\,_{2}F_{1}(a+1-c,b+1-c;2-c;-z)\right]~.
\end{equation}
Now, we can rescale the constant of integration in the above solution and rewrite the solution as
\begin{equation}\label{sol_aa}
    R(z) = z^{\alpha_0}(1+z)^{\beta_0}\left[C_1 \,_{2}F_{1}(a,b;c;-z) + C_2\,z^{1-c}\,_{2}F_{1}(a+1-c,b+1-c;2-c;-z)\right]~.
\end{equation}
If we take one such value for $\alpha_0$ and $\beta_0$ as
\begin{equation}
    \alpha_0 = -i\frac{\omega-m\Omega_+}{2\kappa_+}~, \qquad \beta_0 = -i\frac{\omega-m\Omega_-}{2\kappa_-}~,
\end{equation}
then the solution (\ref{sol_aa}) becomes \ref{solution_1}, which is our desired equation.

\section{Calculations for extremal rotating BTZ black hole}\label{app_2}
In this appendix, We will describe the steps involved in arriving at \ref{eq_ext_BTZ_BH} from \ref{radial_perturbation}. We will also describe the steps for calculating the solution of \ref{eq_ext_BTZ_BH}.

\subsection{Transforming the radial Klein-Gordon equation}\label{app_2a}

The transformed coordinate defined for an extremal rotating BTZ black hole is $\tilde{z}=(r^2-r_+^2)/r_+^2$ (\ref{transformation_1}). If we apply the above transformation,
the radial source-free Klein-Gordon equation (\ref{radial_perturbation}) becomes
\begin{equation}\label{eq_ext_BTZ_BH_a}
    R''(\tilde{z}) + \frac{r_+^2}{2r^2}\left[2+r\frac{f'(r)}{f(r)}\right]R'(\tilde{z})+\frac{r_+^4}{4r^2f(r)}\left[-\frac{m^2}{r^2}+\frac{1}{f(r)}\{\omega-m\Omega(r)\}^2\right]R(\tilde{z})=0~.
\end{equation}
For an extremal rotating BTZ black hole $r_+ = r_-$, i.e, $f(r)$ and $\Omega(r)$ yield (from \ref{def_f_Omega})
\begin{equation}\label{def_f_Omega_1}
    f(r) = \frac{(r^2-r_+^2)^2}{l^2r^2}~,\qquad \Omega(r) = \frac{r_+^2}{lr^2}~.
\end{equation}
It implies
\begin{equation}
    \frac{f'(r)}{f(r)} = \frac{4r}{(r^2-r_+^2)} - \frac{2}{r}~,
\end{equation}
therefore, the coefficient of $R'(\tilde{z})$ can be expressed as:
\begin{equation}\label{coeff_R'(z)_1}
    \frac{r_+^2}{2r^2}\left[2+r\frac{f'(r)}{f(r)}\right] = \frac{2}{\tilde{z}}~.
\end{equation}
Similarly, the coefficient of $R(\tilde{z})$ can be written as:
\begin{equation}
    \frac{r_+^4}{4r^2f(r)}\left[-\frac{m^2}{r^2}+\frac{1}{f(r)}\{\omega-m\Omega(r)\}^2\right] = \frac{l^2}{4\tilde{z}^2}\left[-\frac{m^2}{r^2}+\frac{1}{f(r)}\{\omega-m\Omega(r)\}^2\right]~.
\end{equation}
Using $r^2 = r_+^2(1+\tilde{z})$ and the form of $f(r)$ and $\Omega(r)$ (from \ref{def_f_Omega_1}), the coefficient of $R(\tilde{z})$ becomes
\begin{equation}\label{coeff_R(z)_1}
    \frac{l^4}{4r_+^2\tilde{z}^4}\left(\omega - \frac{m}{l}\right)^2 + \frac{l^4}{4r_+^2\tilde{z}^3}\left(\omega^2 - \frac{m^2}{l^2}\right)~.
\end{equation}
If we define $A$ and $B$ as in \ref{def_A_B},
then the coefficient of $R(\tilde{z})$ can be written as $(A\,B\,\tilde{z} +B^2)/4\tilde{z}^4$~.
Thus, in the transformed coordinate, the radial source-free Klein-Gordon equation (\ref{eq_ext_BTZ_BH_a}) becomes 
\begin{equation}
    R''(\tilde{z}) + \frac{2}{\tilde{z}}R'(\tilde{z})+\frac{1}{4\tilde{z}^4}\left(AB\tilde{z} +B^2\right)R(\tilde{z})=0~,
\end{equation}
which is our desired equation (\ref{eq_ext_BTZ_BH}).

\subsection{Solution of the transformed radial Klein-Gordon equation}\label{app_2b}

In this section, we will describe the steps involved in calculating the solution of \ref{eq_ext_BTZ_BH}.
If we apply the transformation $y= 1/\tilde{z}$, \ref{eq_ext_BTZ_BH} becomes
\begin{equation}\label{transformed_ext__BTZ_BH}
    \frac{\mathrm{d}^2R}{\mathrm{d}y^2}+\frac{1}{4}\left(\frac{AB}{y} +B^2\right)R(y)=0~.
\end{equation}
It can be checked that this equation has a regular singular point at $y=0$ and an irregular singular point of rank $1$ at $y=\infty$. The confluent hypergeometric differential equation also has the same singular points. It means we can transform this equation to the confluent hypergeometric differential equation with the substitution
\begin{equation}
    R(y) = y^\alpha e^{\beta y} G(y)~,
\end{equation}
by choosing the suitable values of $\alpha$ and $\beta$. With the above substitution, the above differential equation (\ref{transformed_ext__BTZ_BH}) becomes
\begin{equation}\label{transformed_ext__BTZ_BH_1}
    y G''(y)+\left(2\alpha+ 2\beta y\right)G'(y)+\left[\frac{\alpha^2-\alpha}{y} + \left(\beta^2 + \frac{1}{4}B^2\right)y+ 2\alpha\beta +\frac{AB}{4}\right]G(y)=0~.
\end{equation}
For confluent hypergeometric differential equation, there should not be $(\alpha^2-\alpha)/y$ and $\left(\beta^2 + B^2/4\right)y$ in the coefficient of $G(y)$. It means
\begin{equation}\label{eq_satisfy_1}
    \alpha^2-\alpha = 0~, \qquad \beta^2 + \frac{1}{4}B^2 =0~.
\end{equation}
If we take any values of $\alpha$ and $\beta$ which satisfy \ref{eq_satisfy_1} say $[\alpha_0, \beta_0]$, then the \ref{transformed_ext__BTZ_BH_1} becomes
\begin{equation}\label{equation_2a}
    y G''(y)+\left(2\alpha_0+ 2\beta_0 y\right)G'(y)+\left[2\alpha_0\beta_0 +\frac{AB}{4}\right]G(y)=0~.
\end{equation}
Let us take $\alpha_0= 0$. In that case, the above differential equation can be written as
\begin{equation}
    y G''(y)+2\beta_0 y G'(y)+\frac{AB}{4}G(y)=0~.
\end{equation}
If we assume $\tilde{x} =-2\beta_0 y$, then the above differential equation transforms to
\begin{equation}
    \tilde{x} G''(\tilde{x}) - \tilde{x} G'(\tilde{x})-\frac{AB}{8\beta_0}G(\tilde{x})=0~.
\end{equation}
On comparing it with the confluent hypergeometric differential equation~\cite{abramowitz_stegun_1972, Arfken_Weber_2005}
\begin{equation}
    \tilde{x}\,G''(\tilde{x}) + (\tilde{c}-\tilde{x})\,G'(\tilde{x}) - \tilde{a}\,G(\tilde{x}) = 0~,
\end{equation}
which has the solution
\begin{equation}
    G(\tilde{x}) = C_1 U(\tilde{a},\tilde{c},\tilde{x}) + C_2\,e^{\tilde{x}} U(\tilde{c}-\tilde{a},\tilde{c},-\tilde{x})~,
\end{equation}
with $C_1$ and $C_2$ as the constants of integration,
we get
\begin{equation}
    \tilde{a} = \frac{AB}{8\beta_0}~, \qquad \tilde{c}=0~.
\end{equation}
Therefore, the solution of \ref{eq_ext_BTZ_BH} is
\begin{equation}\label{sol_ext_rot_BTZ_BH_a1}
    R(\tilde{z}) = e^{\frac{\beta_0}{\tilde{z}}}\left[C_1 U\left(\tilde{a},\tilde{c},-\frac{2\beta_0}{\tilde{z}}\right) + C_2\,e^{-\frac{2\beta_0}{\tilde{z}}} U\left(\tilde{c}-\tilde{a},\tilde{c},\frac{2\beta_0}{\tilde{z}}\right)\right]~.
\end{equation}
If we take $\beta_0 = iB/2$, then the solution (\ref{sol_ext_rot_BTZ_BH_a1}) becomes \ref{ext_BTZ_BH_sol}, which is our desired result.
\section{Calculation for charged BTZ black hole}\label{app_C}
In this appendix, we will chalk out the steps involved in arriving at the approximate form of $f(r)$ and $\Omega(r)$ in \ref{charges_rotating_BTZ_f(r)_approximated} for a charged rotating BTZ black hole.

For reference, we are rewriting the form of $f(r)$ and $\Omega(r)$ for charged rotating BTZ black hole (\ref{form_of_f(r)}):
\begin{equation}
    f(r) = -M + \frac{r^2}{l^2} + \frac{J^2}{4r^2}- 2q^2\ln{\left(\frac{r}{l}\right)} \equiv -M + \frac{r^2}{l^2} + \frac{J^2}{4r^2}- q^2\ln{\left(\frac{r^2}{l^2}\right)}~,\qquad \Omega(r) = \frac{J}{2r^2}~.
\end{equation}
Since we want to write $f(r)$ and $\Omega(r)$ as similar to \ref{def_f_Omega}
, we can apply a transformation $\tilde{z} = (r^2-r_+^2)/r_+^2$, which implies
\begin{equation}
    2q^2\ln{\left(\frac{r}{l}\right)} = q^2\ln{\left[\frac{r_+^2(1+\tilde{z})}{l^2}\right]} \equiv q^2\ln{\left(\frac{r_+^2}{l^2}\right)}+q^2\ln{\left(1+\tilde{z}\right)}.
\end{equation}
In the new coordinate, the event horizon is at $\tilde{z}=0$. For a slowly varying tidal field ($M\omega\ll 1$) and near zone approximation ($M\omega \tilde{z} \ll1$), we can neglect the terms of order $q^2\mathcal{O}(\tilde{z}^2)$ or $\mathcal{O}(M^2\omega^2\tilde{z}^2)$ for a BTZ black hole with small charge ($q\sim \mathcal{O}(M\omega)$). It yields
\begin{equation}
    2q^2\ln{\left(\frac{r}{l}\right)} \sim q^2\ln{\left(\frac{r_+^2}{l^2}\right)}+q^2 \tilde{z} \equiv -q^2 + q^2\ln{\left(\frac{r_+^2}{l^2}\right)}+q^2 \frac{r^2}{r_+^2}.
\end{equation}
Now $f(r)$ becomes
\begin{equation}
    f(r) = -M + \frac{r^2}{l^2} + \frac{J^2}{4r^2}- 2q^2\ln{\left(\frac{r}{l}\right)} \sim -\left[M -q^2+q^2\ln{\left(\frac{r_+^2}{l^2}\right)}\right] + \left(1-\frac{q^2l^2}{r_+^2}\right)\frac{r^2}{l^2}+ \frac{J^2}{4r^2}.
\end{equation}
For brevity, we can re-express it as
\begin{equation}
    f(r) \sim -\tilde{M} + \frac{r^2}{\tilde{l}^2} +\frac{J^2}{4r^2}.
\end{equation}
where the form of $\tilde{M}$ and $\tilde{l}$ are given in \ref{tilde_M_tilde_l}.
\bibliographystyle{apsrev4-1}
\bibliography{ref}
\end{document}